\documentclass[10pt,final]{elsart}
\usepackage{amsmath,amssymb,graphicx,epsfig}

\begin{document}
\begin{frontmatter}
\title{
Variable resistance at the boundary
between semimetal and excitonic insulator}
\author[modena]{Massimo Rontani\corauthref{cor}},
\corauth[cor]{Corresponding author.}
\ead{{\tt rontani@unimore.it}}
\author[ucsd]{L. J. Sham}
\address[modena]{%
INFM National Center on nanoStructures
and bioSystems at Surfaces (S3), \\ Via Campi 213/A, 41100 Modena, Italy}
\address[ucsd]{%
Department of Physics, University of California San Diego, \\
La Jolla, California 92093-0319}
\begin{abstract}
We solve the two-band model for the transport across a junction
between a semimetal and an excitonic insulator.
We analyze the current in terms of two competing terms
associated with neutral excitons and charged carriers, respectively.
We find a high value for the interface resistance, extremely
sensitive to the junction transparency. We 
explore favorable systems for 
experimental confirmation.

\end{abstract}
\end{frontmatter}
\section{Introduction}

The concept that excitons can condense in a semimetal (SM)
and form an {\em excitonic insulator} (EI),
if the energy band overlap is small compared to their binding
energy,
dates back to the sixties \cite{review}.
Experimental evidence has been put forward for the exciton phase
\cite{exp}, but the EI
state remains a mystery. Moreover, the possibility of experimental
discrimination between the EI and the ordinary
dielectric has been called into question \cite{Keldyshmaligno}.
We demonstrate that, {\em if} an EI exists, it develops
unusual transport properties that make it qualitatively
different from an ordinary insulator.

Elsewhere \cite{prl05} we considered, in a two-band model, 
a junction between a SM and a
semiconductor, whose small gap originates from the renormalization of
the SM energy bands due to (i) hybridization of
conduction and valence bands (ii) electron-hole pairing driving
the EI condensation. Carriers incident on the interface from the SM
side with energies below the gap are backscattered
again into the SM, possibly into a different band.
We found that interband scattering only occurs for (ii),
due to the proximity of the EI which broadens the interface potential
profile.

Here we focus on the latter case only.
We analyze the current generated by a bias voltage across
a clean SM/EI junction as two competing terms associated
with neutral excitons and charged carriers, respectively.
Below the EI gap, carriers
are backscattered by the interface with energy band branch crossing.
The formalism is similar to that for the metal/superconductor
(NS) interface \cite{Andreev},
and indeed we find the same dependence of transmission and
reflection coefficients on the quasi-particle energy $\omega$.
However, while electrons below the superconducting
gap are Andreev-reflected as holes,
carriers reflected below the EI gap
conserve their charge and the electric current is
zero.
Above the gap, when charge transmission is allowed,
an unusually high electrical resistance remains.
We find that the electrons that
are backscattered from one band to another
are equivalent to incoming holes correlated with
the incoming electrons. When such pairs enter the
condensate they are converted into an exciton supercurrent,
in such a way that the electron-hole flow across the sample is conserved.
The latter exciton channel
is preferred with respect to charge transmission,
even if $\omega$ is just slightly above the gap.
Therefore, the additional resistance arises due to the competiton
of exciton and charge currents, reminescent
of the interplay between electric supercurrent and
heat flow at the NS junction. The effect is smeared as an
insulating overlayer is inserted at the interface, spoiling
the transparency of the junction: in the tunneling limit,
exciton transport is suppressed.
We further discuss physical systems 
which could show the effects our theory predicts.

The paper is organized as follows: In Sec.~\ref{Andreevm}
we describe 
the 
solution of the electron transmission through the interface in terms 
of the two-band model of the 
SM/EI junction and
in Sec.~\ref{example} we analyze the transport in terms of charge
and exciton currents
and examine the role of the exciton coherence. 
Then we study the interface differential conductance 
(Sec.~\ref{dIdV}), and 
lastly we review candidate experimental systems (Sec.~\ref{exp}).

\section{Transport across the interface}
\label{Andreevm}

We consider a junction made of a semimetal and an excitonic
insulator. Specifically, the EI
band structure originates from the renormalization of the SM energy bands,
driven by Coulomb interaction. The EI gap
corresponds to the binding energy of the
electron-hole pairs which form a condensate. The interface
discontinuity is solely
brought about by the variation of the electron-hole pairing potential,
$\Delta\!\left(z\right)$. This kind of junction
could be experimentally realized by applying a pressure gradient or
by inhomogeneously doping a sample
grown by means of epitaxial techniques (see Sec.~\ref{exp}).

The electron and hole Fermi surfaces of the SM on the
junction left-hand side are taken to be perfectly nested,
the effective masses of the
two bands being isotropic and equal to $m$.
The quasi-particle excitations across the interface must satisfy the
mean-field equations
\begin{subequations}
\label{eq:BdGsimple}
\begin{eqnarray}
\omega \,f\!\left(z\right) &=&
-\frac{1}{2m}\left[\frac{{\partial}^2}{\partial z^2}+k_{\text{F}}^2
\right]f\!\left(z\right)
+\Delta\!\left(z\right) g\!\left(z\right), \\
\omega \,g\!\left(z\right) &=&
\frac{1}{2m}\left[\frac{{\partial}^2}{\partial z^2}+k_{\text{F}}^2\right]
g\!\left(z\right) + \Delta\!\left(z\right) f\!\left(z\right),
\end{eqnarray}
\end{subequations}
with $k_{\text{F}}$ Fermi wave vector and $\hbar=1$.
The amplitudes $f$ and $g$ are the position-space representation
of the electron quasi-particle across the interface: $\left|f\right|^2$
($\left|g\right|^2$) is the probability for an electron of being
in the conduction (valence) band, with energy $\omega>0$ referenced
from the chemical potential, which is in the middle of the EI
gap due to symmetry. We assume $\Delta$ is a step function,
\protect{$\Delta\!\left(z\right) = \Delta\, \theta\!\left(z\right)$}.

In the elastic scattering process at the interface,
all relevant quasi-particle
states are those degenerate --- with energy $\omega$ --- on both sides
of the junction. We handle the interface by matching wave functions
of the incident, transmitted, and reflected particles at the
boundary. In the bulk EI, there are a pair of magnitudes of
$k$ associated with $\omega$, namely
\begin{equation}
k^{\pm}=\sqrt{2m}\sqrt{ k^2_{\text F}/2m \pm
\left(\omega^2 - {\Delta}^2\right)^{1/2} }.
\label{eq:kvect}
\end{equation}
The total degeneracy of relevant states for each $\omega$ is fourfold:
$\pm k^{\pm}$.
The two states $\pm k^+$
have a dominant conduction-band character, while the two states $\pm k^-$
are mainly valence-band states.
Using the notation
\begin{equation}
\Psi(z)=
{f\!\left(z\right) \choose
g\!\left(z\right)}
\end{equation}
the wave functions degenerate in $\omega$ are
\begin{equation}
\Psi_{\pm k^+}=
{ u_0 \choose v_0 } {\rm e}^{ \pm\text{i} k^+z },
\qquad
\Psi_{\pm k^-}=
{ v_0 \choose u_0 } {\rm e}^{ \pm\text{i} k^-z },
\label{eq:bulk1D}
\end{equation}
with the amplitudes $u_0,v_0$ defined as
\begin{equation}
u_0=\sqrt{\frac{1}{2}\left[1+\frac{(\omega^2-{\Delta}^2)^{1/2}}{\omega}
\right]},\quad
v_0=\sqrt{\frac{1}{2}\left[1-\frac{(\omega^2-{\Delta}^2)^{1/2}}{\omega}
\right]},
\end{equation}
possibly extended in the complex manifold.
With regards to the SM bulk,
$\Delta=0$ and the two possible magnitudes of the momentum $q$ reduce to
\protect{$q^{\pm}=[2m( k^2_{\text F}/2m \pm \omega  )]^{1/2}$},
with wave functions
\begin{equation}
\Psi_{\pm q^+}=
{ 1 \choose 0 } {\rm e}^{ \pm\text{i} q^+z },
\qquad
\Psi_{\pm q^-}=
{ 0 \choose 1 } {\rm e}^{ \pm\text{i} q^-z },
\label{eq:SMbulk1D}
\end{equation}
for conduction and valence bands, respectively.

The effect of an insulating layer or of localized disorder at the
interface is modeled by a
$\delta$-function potential, namely $V\!(z)=H\delta\!(z)$.
The appropriate boundary conditions, for particles traveling from
SM to EI are as follows: (i) Continuity of
$\Psi$ at $z=0$, so \protect{$\Psi_{\text{EI}}(0)=\Psi_{\text{SM}}(0)
\equiv \Psi(0)$}. (ii) \protect{$\left[f_{\text{EI}}'(0)-
f_{\text{SM}}'(0)\right]/(2m)=Hf(0)$} and
\protect{$\left[g_{\text{EI}}'(0)-
g_{\text{SM}}'(0)\right]/(2m)=-Hg(0)$}, the derivative boundary
conditions appropriate for $\delta$-functions
\cite{boundary}. (iii) Incoming (incident),
reflected and transmitted wave directions are defined by their group
velocities. We assume the incoming conduction band electron produces
only outgoing particles, namely an electron incident from the left can
only produce transmitted particles with positive group velocities
$v_{\text{g}}>0$ and reflected ones with $v_{\text{g}}<0$.

Consider an electron incident on the interface from the SM with energy
$\omega>\Delta$
and wave vector $q^+$. There are four channels for
outgoing particles, with probabilities $A$, $B$, $C$, $D$,
and wave vectors $q^-$, $-q^+$, $k^+$, $-k^-$, respectively.
In other words, 
$C$ is the probability of transmission 
through the
interface with a wave vector on the same 
(i.e., forward) side of its Fermi surface as $q^+$ 
(i.e., $q^+\rightarrow k^+$, not 
$-k^-$), while $D$ gives the probability
of transmission on the back side of the Fermi surface (i.e., $q^+\rightarrow
-k^-$). $B$ is the probability of intraband reflection, while
$A$ is the probability of reflection on the 
forward side of the Fermi surface (interband
scattering from conduction to valence band).
We write the steady state solution as
\begin{displaymath}
\Psi_{\text{SM}}(z) = \Psi_{\text{inc}}(z)+\Psi_{\text{refl}}(z),
\qquad \Psi_{\text{EI}}(z)=\Psi_{\text{trans}}(z),
\end{displaymath}
where
\begin{eqnarray}
&&\Psi_{\text{inc}}(z) =
{ 1 \choose 0 } {\rm e}^{ \text{i} q^+z }, \quad
\Psi_{\text{refl}}(z)  =
a { 0 \choose 1 } {\rm e}^{ \text{i} q^-z } +
b { 1 \choose 0 } {\rm e}^{ -\text{i} q^+z },\nonumber\\
&&\Psi_{\text{trans}}(z)  =
c { u_0 \choose v_0 } {\rm e}^{ \text{i} k^+z } +
d { v_0 \choose u_0 } {\rm e}^{ -\text{i} k^-z }.
\label{eq:boundary}
\end{eqnarray}
Applying the boundary conditions, we obtain a system of four linear
equations in the four unknowns $a$, $b$, $c$, and $d$, which we solve
at a fixed value for $\omega$. We introduce the
dimensionless barrier strength
\protect{$Z=mH/k_{\text{F}}=H/v_{\text{F}}$}, where 
$v_{\text{F}}$ is the Fermi velocity.
The quantities $A$, $B$, $C$, $D$, are the ratios of the probability
current densities of the specific transmission or reflection channels
to the current of the incident particle,
e.g.~$A=\left|J_A/J_{\text{inc}}\right|$, and so on.
The conservation of probability requires that
\begin{equation}
A + B + C + D = 1.
\label{eq:coeffsum}
\end{equation}
This result is useful in simplifying expressions for energies below
the gap, $\omega<\Delta$, where there can be no transmitted electrons,
so that $C=D=0$. Then, Eq.~(\ref{eq:coeffsum}) reduces simply to
\protect{$A=1-B$}.

\begin{figure}
\centerline{\epsfig{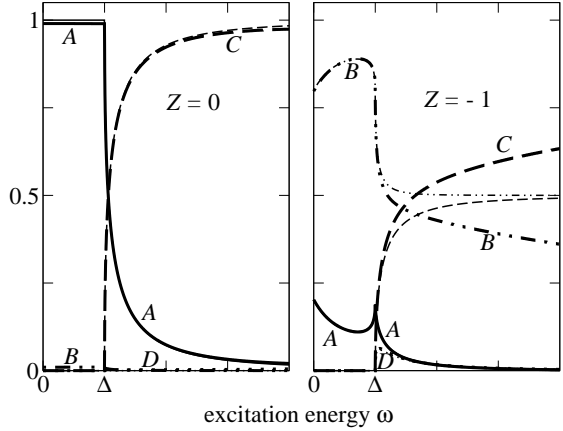}}
\caption{
Plot of transmission and reflection coefficients at SM/EI boundary
computed both in the Andreev approximation (thin lines) and taking
exactly into account the wave vectors of scattered particles (thick lines).
Only in the latter case the coefficients depend on $\Delta/\left|G\right|$
(we take $\Delta/\left|G\right|=0.1$). Left: $Z=0$. Right: $Z=-1$.
$A$ gives the probability of interband reflection, $B$ gives
the probability of ordinary intraband reflection, $C$ gives the transmission
probability without branch crossing, and $D$ gives the probability
of transmission with branch crossing. The parameter $Z$ measures the
interface transparency.
}
\label{fig5}
\end{figure}

The Andreev approximation \cite{Andreev} consists in letting
$k^+=k^-=q^+=q^-=k_{\text{F}}$ in Eqs.~(\ref{eq:boundary}),
on the basis that the ratio $\Delta/\left|G\right|$ is small,
where $G$ is the band overlap of the SM.
Figure \ref{fig5} compares results
obtained in the framework of the Andreev approximation
(thin lines) with data computed without any constraint on
momenta of quasi-particles (thick lines), at
\protect{$\Delta/\left|G\right|=0.1$}.
While the agreement at $Z=0$ is satisfactory, 
the coefficients at $Z=-1$ 
deviate significantly
for energies above the gap.
Note that, whatever the value of $\Delta/\left|G\right|$ or $Z$ is,
the full numerical calculation always gives finite values for $B$ and 
$D$, contrary to the approximate analytic 
results according to which $B=D=0$
when $Z=0$. As the ratio \protect{$\Delta/\left|G\right|$}
increases, the agreement between approximate and full solutions
turns out to be increasingly worse.

If the junction is clean ($Z=0$, left panel of Fig.~\ref{fig5}),
below the gap, $\omega < \Delta$, only interband reflection is possible.
Even above the gap, $\omega > \Delta$, there is a high
probability for interband reflection, which strongly
depends on $\omega$: for energies close to the gap, $\omega \approx \Delta$,
reflection is almost certain, $A\approx 1$. Remarkably,
transmission probability $C \approx 1 - A$ increases very slowly with
$\omega$, which is the cause for the high value of resistance.
The effect is washed out by the opacity of the interface: as
$\left|Z\right|$ increases ($Z=-1$, right panel of Fig.~\ref{fig5}),
transmission probability loses its dependence on $\omega$, and reflection
channel turns from interband, $A$, into intra-band type, $B$.

Results obtained for the SM/EI
junction by means of the Andreev approximation
are formally identical to those of the NS interface,
as given in Table II of Ref.~\cite{BTK}.
However, there are a few differences in the dependence of the NS
coefficients on $Z$ with respect to the present case, which is due to
different boundary conditions, as stressed in note \cite{boundary}.
While the NS coefficients are even functions of $Z$, in the SM/EI case
$A$ and $B$ do not have a definite parity with respect to the sign of $Z$
for $\omega<\Delta$,
while for $\omega>\Delta$, $A$, $B$, $C$, and $D$ are even in $Z$.
Nevertheless, the expressions for coefficients in the strong barrier case
coincide with the corresponding ones for the NS case.
Therefore, apart from some differences for small values of $Z$,
the physical role of the barrier is the same in both cases.

\section{\label{example} Charge versus exciton current}

We describe the interband reflection process in terms of a
neutral electron-hole current.
The probability density
$\rho_{\text{e-h}}\!\left(z,t\right)$ for finding either a conduction-band
electron or a valence-band hole at a particular time and place is
\protect{$\rho_{\text{e-h}}\!\left(z,t\right)=
\left|f\right|^2 + 1 - \left|g\right|^2$}.
We consider conduction electrons with crystal
momentum with modulus larger than
$k_{\text{F}}$, otherwise we define
$\rho_{\text{e-h}}\!\left(z,t\right)$
as \protect{$\rho_{\text{e-h}}\!\left(z,t\right)=
1 - \left|f\right|^2 + \left|g\right|^2$}.
We obtain, in the first case,
\begin{equation}
\frac{\partial \rho_{\text{e-h}}}{\partial t}+ \frac{\partial
J_{\text{e-h}}}{\partial z}=0, \qquad
J_{\text{e-h}} = J_{\text{pair}} + J_{\text{cond}},
\label{eq:continuity2}
\end{equation}
where \protect{$J_{\text{pair}}=m^{-1}{\text{Im}}\{
f^*\partial f / \partial z + g^*\partial g / \partial z\}$} is the
density current of the electron-hole pair,
and the term \protect{$ \partial J_{\text{cond}}/\partial z =
-4\,{\text{Im}}\{ f^* g \, \Delta \}$}
explicitly depends on
the built-in coherence of the electron-hole condensate $\Delta$.
While $J_{\text{pair}}$ is analogous to the standard
particle current \protect{$J=m^{-1}{\text{Im}}\{
f^*\partial f/\partial z - g^*\partial g / \partial z \}$}
except a difference in sign, the
term $J_{\text{cond}}$ is qualitatively different and
is attributed to the exciton supercurrent of the EI ground state.

If $\omega < \Delta$ and $Z=0$, each wave function (\ref{eq:boundary}),
solution of Eqs.~(\ref{eq:BdGsimple}),
carries zero total electric current $eJ$, which is the
sum of the equal and opposite
incident and reflected fluxes, and finite and constant
electron-hole current \protect{$J_{\text{e-h}}=2v_{\text{F}}$}.
Inside the SM side ($z<0$), the supercurrent contribution
$J_{\text{cond}}$ is zero. Note that
\protect{$J_{\text{e-h}}$} conserves
its constant value, independent of $z$, since quasi-particle states
(\ref{eq:boundary}) are stationary.
In fact, as the contribution to the electron-hole current
$J_{\text{pair}}$
vanishes approaching the boundary, $J_{\text{pair}}$ is
rapidly converted into the supercurrent $J_{\text{cond}}$.
Excitons therefore can flow into the EI side without any resistance,
and the sum \protect{$J_{\text{e-h}}$} of the two contributions,
$J_{\text{pair}}$ and $J_{\text{cond}}$,
is constant through all the space.

As an example,
consider the quasi-particle steady
state of Eq.~(\ref{eq:boundary}) and
the coefficients $a$, $b$, $c$, $d$ obtained in the
``Andreev approximation'' (Sec.~\ref{Andreevm}).
For $\omega<\Delta$, $k^+$ and $k^-$ in the excitonic
insulator have small imaginary part which lead to an exponential decay on
a length
scale $\lambda$, where
\begin{equation}
\lambda = \frac{v_{\text{F}}}{2\Delta}\left(1-\frac{\omega^2}{\Delta^2}
\right)^{-1/2}.
\end{equation}
The quasi-particles penetrate a depth $\lambda$ before
the electron-hole current $J_{\text{pair}}$ is converted to
a supercurrent $J_{\text{cond}}$ carried by the condensate; right
at the gap edge the length diverges. For clarity, we define
$C$ and $D$ here as the transmission probabilities at $z\gg \lambda$,
while for $\omega>\Delta$ plane-wave currents are spatially uniform
and we need not specify the position at which they are evaluated.

When there is no barrier at the interface, $Z=0$, the steady state
(\ref{eq:boundary}) is specified by $b=d=0$, $a=v_0/u_0$,
and $c=1/u_0$. Below the gap
coherence factors $u_0$ and $v_0$ are complex and
equal in modulus. For $\omega<\Delta$, $\left|a\right|^2=1$, which
means the incident conduction-band electron is totally
reflected into the SM valence band. Thus, the electron-hole
current  $J_{\text{pair}}$ carried in the semimetal equals
$2v_{\text{F}}$, but $J_{\text{pair}}$ of the excitonic insulator
is exponentially small for $z\gg 0$. Explicitly,
\begin{displaymath}
J_{\text{pair}} = \frac{\left|c\right|^2}{m}(\left|u_0\right|^2+
\left|v_0\right|^2)\,{\text{Im}}\!\left[({\rm e}^{ \text{i} k^+z }
)^*\frac{\partial}{\partial z} ({\rm e}^{ \text{i} k^+z })\right].
\end{displaymath}
Letting \protect{$k^+\approx k_{\text{F}} + {\text{i}}/(2\lambda)$},
we have
\begin{equation}
J_{\text{pair}} = 2v_{\text{F}}{\rm e}^{-z/\lambda}.
\end{equation}
The ``disappearing'' electron-hole current reappears as
exciton current carried by the condensate. Recalling the
definition of $J_{\text{cond}}$,
\begin{displaymath}
\partial J_{\text{cond}} / \partial z =
-4\,{\text{Im}}\!\left\{ f^* g \, \Delta \right\},
\end{displaymath}
by integration we obtain
\begin{equation}
{J}_{\text{cond}} = -4\Delta\left|c\right|^2\int_0^z\!\!{\text{d}}\,z'
{\rm e}^{-z'/\lambda}\,{\text{Im}}\!\left[u_0^*v_0\right]
= 2v_{\text{F}}\left(1-{\rm e}^{-z/\lambda}\right).
\end{equation}
This is the desired result, explicitly showing the supercurrent
${J}_{\text{cond}}$
increasing to an asymptotic value as $z\rightarrow \infty$, at the
same rate as the quasi-particle current $J_{\text{pair}}$ dies away.

Above the gap, $\omega> \Delta$, $J$ increases from zero
and \protect{$J_{\text{e-h}}$} decreases.
However, close to the gap, electron transmission to the EI side is
still inhibited ($C\approx 0$) by the pairing between
electrons and holes of the condensate: an electron can stand alone and
carry current only after its parent exciton has been ``ionized''
by injecting --- say ---  a conduction-band electron or by filling
a valence-band hole in the EI. The ionization costs
an amount of energy of the order of the binding energy of the
exciton, $\Delta$. Therefore, as long as $\omega \approx \Delta$,
the competition between exciton and electron flow favors interband
reflection, which is the source of the high electric resistance.

\section{Differential conductance at finite voltage}\label{dIdV}

Electric transport across the SM/EI interface is the experimental signature
of the physics we have previously discussed. When a bias voltage $V$ is
applied across the junction, nonequilibrium quasi-particle populations
are generated, which
can be found in principle only by implementing a self-consistent scheme
linking the computation of both charge and potential. Here we adopt
a simplified approach assuming ballistic acceleration of particles
except for
the scattering at the interface. This should be a good
approximation for the case e.g.~of a 
thin junction connecting massive electrodes,
as long as the diameter of the orifice is small compared to a mean-free
path. In addition, we assume that the distribution functions of all
incoming particles are given by equilibrium Fermi functions, apart from
the energy shift due to the accelerating potential. We choose the
electrochemical potential in the EI as our reference level, being a
well defined quantity at finite temperature $T$, when carriers
are provided by thermal excitations.

The computation of the electric current $I$ follows step by step the
analogous treatment 
in the superconductor case \cite{BTK,usunpublished}. Here we only
state the result for the differential conductance,
\protect{$\partial I / \partial V$}, which in ordinary units is
\begin{equation}
\frac{\partial I}{\partial V} = e^2WN\!(\varepsilon_{\text{F}})
\frac{v_{\text{F}}}{4}
\int_{-\infty}^{\infty}
\!\!\! \text{d}\,\omega\,
\xi (\omega) \left[C(\omega) + D(\omega) \right]
\left[-\frac{\partial f(\omega ')}{\partial \omega '}
\right]_{\omega ' = \omega - eV},
\label{eq:dIdV}
\end{equation}
where $C(\omega)$ and $D(\omega)$ are taken to be even functions
defined over the whole real axis, $f(\omega)$ is the Fermi distribution
function, $\xi (\omega)$  is the channel degeneracy which takes the value
one (two) if
$\left|\omega \right| > \left|G\right|/2$ ($\left|\omega \right| <
\left|G\right|/2$), $W$ is the interface cross-sectional area,
and $N\!(\varepsilon_{\text{F}})$ is the density of states per volume
at the Fermi energy per each semimetal band.
Equation~(\ref{eq:dIdV}) is derived assuming that the transmission
coefficients are independent of $V$.
At $T=0$ the function
\protect{$-\partial f / \partial \omega$} appearing in
Eq.~(\ref{eq:dIdV}) turns into a Dirac's delta, while at finite
$T$ one must perform the integration over $\omega$ and the overall
effect is that sharp energy features of
\protect{$\partial I / \partial V$} are smeared out.
We focus exclusively on the zero temperature case.

\begin{figure}
\centerline{\epsfig{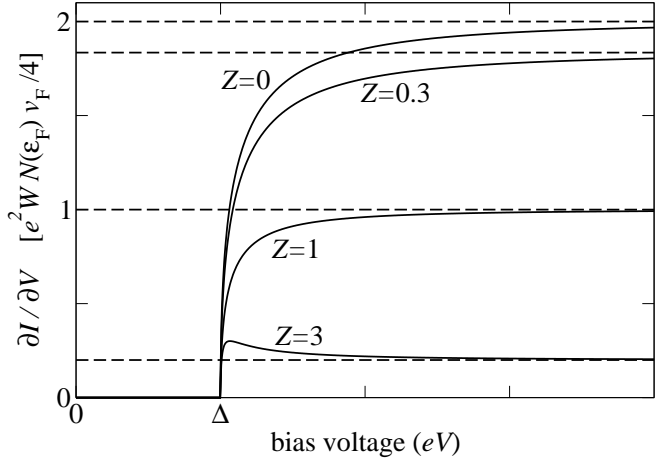}}
\caption{
Plot of differential conductance,
\protect{$\partial I / \partial V$}, computed at zero
temperature in the Andreev approximation,
as a function of the bias voltage applied at SM/EI boundary for several
values of the barrier transparency $Z$.
Curves for different values of $Z$ at large voltages tend to
asymptotic values (dashed lines) given by the contact resistance of the
junction in the absence of the electron-hole condensate ($\Delta=0$).
The differential
conductance is given in units of $e^2WN\!(\varepsilon_{\text{F}})
v_{\text{F}}/4$, where $W$ is the interface cross-sectional area,
$v_{\text{F}}$ is the Fermi velocity,
and $N\!(\varepsilon_{\text{F}})$ is the density of states per volume
at the Fermi energy per each semimetal band.
}
\label{fig7}
\end{figure}

Figure \ref{fig7} shows the differential conductance
\protect{$\partial I / \partial V$} of the SM/EI interface
at $T=0$ as a function
of the bias voltage and for different values of $Z$. The calculation
has been carried out in the Andreev approximation.
The current shows an activated behavior, the
threshold being the energy gap $\Delta$. When the interface is clean
($Z=0$), the conductance slowly rises with the voltage $V$,
due to the additional
resistance brought about by the interband reflection mechanism. In fact,
\protect{$\partial I / \partial V$} goes like
\protect{$(\left|eV\right|-\Delta)^{1/2}$},
as discussed in Sec.~\ref{Andreevm}.
As the interface opacity gradually increases (going from $Z=0.3$ up to
$Z=3$) we note the following two features: (i) Curves become
progressively more and more flat, with a well defined step
at the threshold $\Delta$. Therefore, the additional resistance close
to the gap, which is responsible for the gradual increase of
\protect{$\partial I / \partial V$}, is completely suppressed in the
tunneling regime. The opacity of the interface spoils
the spatial coherence between the SM and EI sides
and inhibites the transport channel $A$. (ii) All curves tend asymptotically,
for large voltages, to a limiting value which is the contact
resistance of the interface when there is no electron-hole
condensate present ($\Delta=0$), namely
\protect{$\partial I / \partial V =
e^2WN\!(\varepsilon_{\text{F}})
v_{\text{F}}
  /2(1+Z^2)$}.
Indeed, at high energies
the effect of the electron-hole condensate is negligible --- while it is
dominant close to $\Delta$ ---: as $Z$ increases, the contact resistance
decreases as $Z^{-2}$ (see the expressions for $C$ and $D$ coefficients
for large values of $Z$ in Table II of Ref.~\cite{BTK}).

\section{Choice of physical systems}\label{exp}

We address the question of
which systems should be considered for the
experimental realization of the SM/EI junction.
The physical quantity which we suggest to measure
is the junction electrical resistance,
in particular the differential conductance
as a function of the applied voltage. We showed this quantity, at
$T=0$ and for different amounts of interface disorder, in Fig.~\ref{fig7}.
By measuring the current we {\em indirectly} probe the effect
of the neutral exciton supercurrent, which is responsible for the loss
of conductance at voltages close to the gap. In such an experiment
it would be important to track the evolution of
conductance as disorder is added to the interface.

\subsection{Rare-earth calcogenides}

Presently, the strongest experimental evidence of
the existence of the EI phase concerns rare-earth
calcogenides such as TmSe$_x$Te$_{1-x}$ \cite{exp},
Sm$_{1-x}$La$_x$S \cite{exp,Wachter2}, Sm$_{1-x}$Tm$_x$S,
YbO and YbS \cite{Wachter2}. These intermediate valent
compounds all crystallize in the NaCl structure and
undergo a semimetal/semiconductor transition under pressure,
since the band overlap $G$ can be changed 
from negative to positive values by applying 
high hydrostatic pressure to the sample, while the dielectric
screening does not change dramatically, since the gap
is indirect \cite{exp}. According to
resistivity and Hall mobility measurements \cite{exp},
at low temperatures one
intercepts the EI phase close to $G\approx 0$. Here we focus
on the most studied TmSe$_{0.45}$Te$_{0.55}$ alloy, but the discussion
could apply to other compounds as well.

When the gap of TmSe$_{0.45}$Te$_{0.55}$ is closing with
external pressure, an indirect band gap
develops between the highest valence Tm 4$f^{13}$ level $\Gamma_{15}$ at
the $\Gamma$ point and the mimimum of the $\Delta_{2^{\prime}}$
conduction band 5$d$ states at the X point of the Brillouin zone \cite{exp}.
Since the otherwise localized 4$f$ band is broadened and shows
a maximum at $\Gamma$ due to $p$(Se,Te)-$f$(Tm) covalent
hybridization \cite{Jansen}, we suggest to realize a SM/EI interface
by varying the hydrostatic pressure applied to a TmSe$_{0.45}$Te$_{0.55}$
sample along the [100] direction. Temperature and pressure
values at which the junction could operate are easily
deduced from the phase diagram shown in Fig.~1 of
Ref.~\cite{superthermal}. For example, a pressure of 14 Kbar
guarantees that the compound remains semimetallic from 5 to 300 K,
while a slight decrease in pressure enters the EI phase
at low temperatures.

\subsection{Vertical transport in layered graphite}

A single planar sheet of graphite is a zero-overlap semimetal. 
Conduction and valence band energy surfaces,
in the proximity of the Fermi energy,
form specular cones whose apexes touch in the two
inequivalent points K and K$^{\prime}$, located at the corners of the
hexagonal two-dimensional Brillouin zone.
These essential-degeneracy points map into each
other by a rotation of $2\pi/6$ \cite{Bassani}.
Interestingly, Coulomb interaction is long ranged due
to the lack of conventional screening \cite{Khveshchenko}.
Khveshchenko \cite{Khveshchenko} claims that graphite hides
a latent excitonic insulator instability.
According to Ref.~\cite{Khveshchenko}, the ground state could be
a charge density wave alternating between
the two inequivalent triangular sublattices, its characteristic wave
vector in reciprocal space connecting K and K$^{\prime}$. A stack
of graphite layers in a staggered (ABAB...) configuration,
with the atoms located in the centers and corners of the hexagons
in two adjacent layers, respectively, could stabilize the EI phase
by enforcing interlayer Coulomb interaction. Also doping could strengthen
the EI ground state inducing exciton ferromagnetism \cite{EIferro}.
This theory seems to explain magnetic correlations recently measured
in highly oriented pyrolitic graphite \cite{Kopelevich}.

We observe that in common layered samples
with AB stacking graphite is a finite-overlap semimetal with very low
carrier concentration, due to the small interlayer tunneling
\cite{Dresselhaus}. The high-symmetry P line connecting K and H
points on the border vertical edge of the three-dimensional Brillouin zone
has still two-fold degeneracy in energy for symmetry
reasons \cite{Jones}, but, due to small band dispersion driven
by interlayer coupling, there is a closed Fermi surface around K
and a hole pocket centered at H. By moving along P one crosses both
electron and hole pockets: the two-dimensional case is recovered
when the interlayer distance increases indefinitely, namely
H coincides with K. Therefore, we propose to fabricate a
SM/EI graphite-based junction by arranging a stacking sequence
where doping or interlayer interaction can be artificially
controlled. Transport occurs in the stacking vertical direction:
the bottom of the relevant conduction band on the SM side of
the junction is located at K point, while the top of valence band at
H$^{\prime}$, where H$^{\prime}$ lies on the P$^{\prime}$ line
including the inequivalent point K$^{\prime}$.

\subsection{Lateral junction of coupled quantum wells}

\begin{figure}
\setlength{\unitlength}{1 cm}
\begin{picture}(18,4)
\put(0.0,-4.0){\epsfig{file=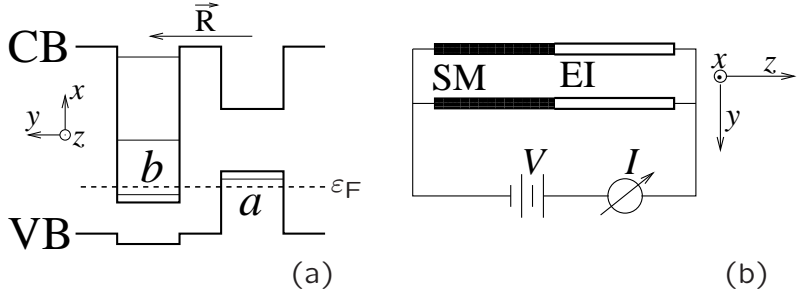,width=3.7in,,angle=90}}
\end{picture}
\caption{
(a) Energy band scheme along the growth direction
of a typical semimetal bilayer
heterostructure. $a$ and $b$ label the highest-energy
valence sub-band in one layer and the lowest-energy
conduction sub-band in the other layer, respectively.
The motion is confined in the $y$ (growth) direction, and it
is two dimensional in the $xz$ plane. The SM/EI interface
lies in the $xy$ plane, and $\bf R$ is the distance
vector between the two layers in real space. (b)
Experimental setup to measure the interface electrical resistance.
A small bias voltage $V$ is laterally applied to both layers
forming the semimetal / excitonic insulator interface, and
an electric current $I$ flows across the interface.
}
\label{figQW}
\end{figure}

Bilayers where electrons and holes are spatially separated
constitue very interesting systems to test ideas presented
in this work, since exciton condensation appears to have been 
observed in these systems \cite{butov}. 
In coupled quantum well heterostructures
a quasi two dimensional semimetal can be realized such that the negative
gap $G$ is indirect in {\em real} space, the valence band edge
in one layer being higher in energy than the conduction band
bottom in the other layer [Fig.~\ref{figQW}(a)]. Below we explain how
our theory can be extened to bilayers in a straightforward
way. Several experimental setups have
been proposed in order to achieve exciton condensation in
such systems, including In$_{1-x}$Ga$_x$As /
AlSb / GaSb$_{1-y}$As$_y$ Type-IIB and
biased modulation-doped GaAs / AlGaAs coupled quantum wells,
and doping ($n$-$i$-$p$-$i$) superlattices
\cite{Datta,Zhu,Naveh,Cheng,Marlow}. There are several
advantages in this scheme. One is that it is possible to
enhance the exciton binding energy by both quantum confinement
and minimization of interlayer tunneling \cite{Datta,Zhu}.
The latter is most conveniently realized by interposing
a wide-gap layer acting as a barrier between the two quantum wells:
the thinner the layer, the stronger the Coulomb electron-hole
attraction. Tunneling must be inhibited to reduce interband virtual
transitions that increase the screening of Coulomb interaction,
which can be accomplished by increasing the height of the
inter-well potential barrier \cite{Datta}. Another key point
is that the semiconductor-to-semimetal transition
can be driven either by manipulating the layer thickness and
material composition or by continuously tuning an external
electric field applied along the growth direction \cite{Naveh}.
Last but not least, high mobility and
low carrier density in state-of-the-art heterostructures
are certainly favorable toward exciton condensation \cite{Marlow}.

We propose to fabricate a lateral SM/EI junction starting from
coupled quantum wells (Fig.~\ref{figQW}). Conduction and valence band
electrons laterally move in the $xz$ plane in spatially separated
quantum wells, while
the interface plane $xy$ extends
parallel to the growth direction (see Fig.~\ref{figQW}). Contrary to
the model of Sec.~\ref{Andreevm}, where conduction- ($b$) or
valence-band ($a$) electrons can overlap in direct space,
different bands imply now spatial separation.
Therefore, the origin of the position vector $\bf r$
for the $b$-electron in one layer now is shifted by the
amount $\bf R$ with respect to position of the $a$-electron
in the other layer [see Fig.~\ref{figQW}(a)].
Besides, the role of an interband hybridization potential
$V_{\text{hyb}}(\bf{k})$ is now played by the hopping
matrix element connecting the two layers via
tunneling. Taking further into account that
the motion is quasi two dimensional (see Ref.~\cite{Naveh}
for the appearance of structure factors in the effective
Coulomb interaction term), equations of motion
(\ref{eq:BdGsimple}) for quasiparticles still hold. The junction
could be realized starting from a coupled quantum well
where exciton condensation has been supposedly achieved
and then destroying pairing in one region of the sample. A method
could be e.g.~to apply a local external electric field along the growth
direction to increase band overlap \cite{Naveh}
and therefore dielectric screening
in order to suppress $\Delta$. Electrodes should
allow to apply a small bias voltage along the lateral direction
[Fig.~\ref{figQW}(b)].
In addition to the interface resistance measurement, this setting
nicely allows for comparison between
the effects of electron-hole
pairing and those of band hybridization, which have been the object of a
recent controversy in cyclotron resonance experiments
\cite{Cheng,Marlow}. In particular, the resistance
measurement we propose is able to elucidate the nature
of the gap that forms in a nominally semimetallic material.

\section{Acknowledgements}
This work is supported by MIUR Progetto
Giovani Ricercatori and MIUR FIRB-RBAU01ZEML.
M.~R.~thanks E.~Randon, E.~K.~Chang, B.~I.~Halperin, and C.~Tejedor
for stimulating discussions.


\begin{thebibliography}{90}
\bibitem{review}
For a review: see B.~I.~Halperin and T.~M.~Rice, Solid State Phys.~21
(1968) 115; S.~A.~Moskalenko and D.~Snoke, Bose Einstein condensation
of excitons and biexcitons, Cambridge University Press,
Cambridge, 2001, Sect.~10.3.
\bibitem{exp} B.~Bucher, P.~Steiner, and P.~Wachter,
Phys.~Rev.~Lett.~67 (1991) 2717; P.~Wachter, A.~Jung,
and P.~Steiner, Phys.~Rev.~B 51 (1995) 5542.
\bibitem{Keldyshmaligno}
R.~R.~Guse\u{\i}nov and L.~V.~Keldysh, Zh.~Eksp.~Teor.~Fiz.~63
(1972) 2255 [English transl.: Soviet Phys.\textendash JETP
36 (1973) 1193].
\bibitem{prl05}
Massimo Rontani and L.~J.~Sham, Phys.~Rev.~Lett.~94 (2005) 186404.
\bibitem{Andreev}
A.~F.~Andreev, Zh.~Eksperim.~Teor.~Fiz.~46 (1964) 1823
[English transl.: Soviet Phys.\textendash JETP 19 (1964) 1228].
\bibitem{boundary}
The latter boundary condition differs in sign from the one appropriate to
the metal/superconductor junction, as outlined in Ref.~\cite{BTK}.
\bibitem{BTK}
G.~E.~Blonder, M.~Tinkham, and T.~M.~Klapwijk, Phys.~Rev.~B
25 (1982) 4515.
\bibitem{usunpublished}
Massimo Rontani and L.~J.~Sham, Appl.~Phys.~Lett.~77
(2000) 3033; cond-mat/0309687.
\bibitem{Wachter2}
P.~Wachter, J.~Alloys and Compounds 225 (1995) 133.
\bibitem{superthermal}
P.~Wachter, B.~Bucher, and J.~Malar, Europhys.~Lett.~62 (2003) 343.
\bibitem{Jansen}
H.~J.~F.~Jansen, A.~J.~Freeman, and R.~Monnier, Phys.~Rev.~B 31
(1985) 4092.
\bibitem{Bassani}
F.~Bassani and G.~Pastori Parravicini, Electronic states
and optical transitions in solids, Pergamon Press, Oxford, 1975.
\bibitem{Khveshchenko}
D.~V.~Khveshchenko, Phys.~Rev.~Lett.~87 (2001) 246802.
\bibitem{EIferro}
B.~A.~Volkov, Yu.~V.~Kopaev, and A.~I.~Rusinov,
Zh.~Eksp.~Teor.~Fiz.~68 (1975) 1899 [English transl.:
Soviet Phys.\textendash JETP 41 (1976) 952];
E.~Bascones, A.~A.~Burkov, and A.~H.~MacDonald,
Phys.~Rev.~Lett.~89 (2002) 86401.
\bibitem{Kopelevich}
Y.~Kopelevich, P.~Esquinazi, J.~H.~S.~Torres, and
S.~Moehlecke, J.~Low Temp.~Phys.~119 (2000) 691.
\bibitem{Dresselhaus}
M.~S.~Dresselhaus and G.~Dresselhaus, Adv.~Phys.~30 (1981) 139.
\bibitem{Jones}
W.~Jones and N.~H.~March, Theoretical solid state physics,
Dover, New York, 1973, Vol.~2, App.~G12.
 \bibitem{butov} 
L.~V.~Butov, C.~W.~Lai, A.~L.~Ivanov, A.~C.~Gossard, and D.~S.~Chemla, 
Nature 417 (2002) 47.
\bibitem{Datta}
S.~Datta, M.~R.~Melloch, and R.~L.~Gunshor, Phys.~Rev.~B 32
(1985) 2607.
\bibitem{Zhu}
X.~Zhu, J.~J.~Quinn, and G.~Gumbs, Solid State Commun.~75
(1990) 595;
X.~Xia, X.~M.~Chen, and J.~J.~Quinn, Phys.~Rev.~B 46 (1992) 7212.
\bibitem{Naveh}
Y.~Naveh and B.~Laikhtman, Appl.~Phys.~Lett.~66 (1995) 1980;
Phys.~Rev.~Lett.~77 (1996) 900.
\bibitem{Cheng}
J.-P.~Cheng, J.~Kono, B.~D.~McCombe, I.~Lo, W.~C.~Mitchel,
and C.~E.~Stutz, Phys.~Rev.~Lett.~74 (1995) 450.
\bibitem{Marlow}
L.~J.~Cooper, N.~K.~Patel, V.~Drouot, E.~H.~Linfield, D.~A.~Ritchie,
and M.~Pepper, Phys.~Rev.~B 57 (1998) 11915;
T.~P.~Marlow, L.~J.~Cooper, N.~K.~Patel, D.~M.~Whittaker,
E.~H.~Linfield, D.~A.~Ritchie, and M.~Pepper, Phys.~Rev.~Lett.~82
(1999) 2362.
\end{thebibliography}
\end{document}